\documentstyle[aps,floats,epsf,euscript]{revtex}

\def\nslash{\rlap{\hspace{0.02cm}/}{n}}
\def\vslash{\rlap{\hspace{0.02cm}/}{v}}
\def\F{{\EuScript F}}
\def\T{{\EuScript T}}

\begin{document}
\draft
\twocolumn[\hsize\textwidth\columnwidth\hsize\csname@twocolumnfalse%
\endcsname

\flushright{CLNS~03/1822\\
{\tt hep-ph/0303082}\\
\vspace{0.2cm}}

\title{Renormalization-Group Evolution of the B-Meson\\
Light-Cone Distribution Amplitude}

\author{Bj\"orn O.\ Lange and Matthias Neubert}

\address{Newman Laboratory for Elementary-Particle Physics, 
Cornell University, Ithaca, NY 14853, USA}
\maketitle

\begin{abstract}
An integro-differential equation governing the evolution of the 
leading-order $B$-meson light-cone distribution amplitude is derived. The 
anomalous dimension in this equation contains a logarithm of the 
renormalization scale, whose coefficient is identified with the cusp 
anomalous dimension of Wilson loops. The exact analytic solution of the 
evolution equation is obtained, from which the asymptotic behavior of the 
distribution amplitude is derived. These results can be used to resum 
Sudakov logarithms entering the hard-scattering kernels in QCD 
factorization theorems for exclusive $B$ decays.
\end{abstract}

\pacs{12.38.Cy,12.39.Hg,12.39.St,13.25.Hw}]
\narrowtext

{\em 1. Introduction.}\quad
It has recently become apparent that the light-cone structure of the $B$ 
meson is of great phenomenological interest. Using standard perturbative 
QCD methods for hard exclusive processes, $B$-meson light-cone 
distribution amplitudes (LCDAs) were first introduced in a study of the 
asymptotic behavior of heavy-meson form factors at large momentum 
transfer \cite{Grozin:1996pq}. Two such amplitudes, called 
$\phi_\pm^B(\omega,\mu)$, arise in the parameterization of $B$-meson 
matrix elements of bilocal heavy-light current operators at leading-order 
in heavy-quark effective theory (HQET). The interest in these quantities 
was revived when it was found that the amplitudes for many exclusive 
$B$-meson decays can be simplified significantly using QCD factorization 
theorems \cite{Beneke:1999br}. At leading power in 
$\Lambda_{\rm QCD}/m_b$ these amplitudes can be expressed in terms of 
convolution integrals of perturbative hard-scattering kernels with the 
leading-order LCDAs. 

The $B$-meson LCDAs appear in processes where light energetic particles 
are emitted into the final state, such as $B\to\pi\pi$ and 
$B\to K^*\gamma$, and at leading power arise from hard interactions of 
collinear partons with the soft spectator quark inside the $B$ meson. In 
almost all applications of QCD factorization theorems only the function 
$\phi_+^B(\omega,\mu)$ contributes. Generically, the hard-spectator term 
in a QCD factorization formula is of the form
\begin{equation}\label{conv}
   \int_0^\infty\frac{d\omega}{\omega}\,T(\omega,\mu)\,
   \phi_+^B(\omega,\mu) \,,
\end{equation}
where the kernel $T$ is calculable in perturbation theory. The factor 
$1/\omega$ ensures convergence of the integral for large $\omega$. The 
characteristic scale for soft-collinear interactions is 
$2E\omega\sim m_b\Lambda_{\rm QCD}$, where $E\sim m_b$ is the energy of 
a collinear particle in the $B$-meson rest frame. At fixed order in 
perturbation theory the kernel depends logarithmically on this scale, 
and it is independent of $\omega$ to lowest order in perturbation 
theory. In addition, the kernel can depend on hard scales of order 
$m_b$. We will assume below that $T$ is defined such that the convolution 
integral (\ref{conv}) is renormalization-group (RG) invariant.

A thorough understanding of factorization requires controlling the scale 
dependence of the LCDA and the kernel under the convolution integral 
(\ref{conv}) using evolution equations. This is crucial for a clean 
separation of physics associated with different mass scales and for the 
systematic resummation of large (Sudakov) logarithms, which enter the 
kernel at every order in perturbation theory. In this Letter we derive 
the RG equations for the LCDA $\phi_+^B(\omega,\mu)$ and for the kernel 
$T(\omega,\mu)$, present their exact analytic solutions, and extract
model-independent results for the asymptotic behavior for $\omega\to 0$ 
and $\omega\to\infty$.

\vspace{0.3cm}
{\em 2. Evolution equations.}\quad
The LCDA is given by the Fourier transform
\begin{equation}
   \phi_+^B(\omega,\mu) = \frac{1}{2\pi} \int d\tau\,e^{i\omega\tau}\,
   \widetilde\phi_+^B(\tau,\mu)
\end{equation}
of a function $\widetilde\phi_+^B(\tau,\mu)$ defined in terms of a 
$B$-meson matrix element in HQET. Denoting by $h$ the effective 
heavy-quark field and by $q_s$ the soft spectator quark, and using a mass 
independent normalization of meson states, we write 
\cite{Grozin:1996pq}
\begin{eqnarray}\label{LCDA}
   &&\langle\,0\,|\,\bar q_s(z)\,S_n(z,0)\,\nslash\,\Gamma\,h(0)\,
    |\bar B(v)\rangle \nonumber\\[0.3cm]
   &&\qquad = - \frac{iF(\mu)}{2}\,\widetilde\phi_+^B(\tau,\mu)\,
    \mbox{tr}\left( \nslash\,\Gamma\,\frac{1+\vslash}{2}\,\gamma_5
    \right) .
\end{eqnarray}
Here $z\parallel n$ with $n^2=0$ is a null vector, $v$ is the $B$-meson 
velocity, $\Gamma$ represents an arbitrary Dirac matrix, and 
$\tau=v\cdot z-i0$. The gauge string $S_n(z,0)$ represents a soft Wilson 
line connecting the points $0$ and $z$ on a straight light-like segment. 
The quantity $F(\mu)$ is the HQET matrix element corresponding to the 
asymptotic value of the product $f_B\sqrt{m_B}$ in the heavy-quark limit.
The distribution amplitude $\widetilde\phi_+^B(\tau,\mu)$ is normalized 
to 1 at $\tau=0$. The analytic properties of this function in the complex 
$\tau$ plane imply that $\phi_+^B(\omega,\mu)=0$ if $\omega<0$.

We denote by $O_+(\omega)$ the Fourier transform of the bilocal HQET
operator in (\ref{LCDA}) and write the relation between bare and 
renormalized operators in the form
\begin{equation}\label{Zdef}
   O_+^{\rm ren}(\omega,\mu) = \int d\omega'\,Z_+(\omega,\omega',\mu)\,
   O_+^{\rm bare}(\omega') \,,
\end{equation}
where $Z_+(\omega,\omega',\mu)=\delta(\omega-\omega')$ at lowest order.
Operators with different momentum $\omega'$ can mix under renormalization 
since they have the same quantum numbers. In the $\overline{\rm MS}$ 
scheme the function $Z_+(\omega,\omega',\mu)$ is defined so as to 
subtract the ultraviolet (UV) poles in the matrix elements of the bare 
operators. 

The $B$-meson matrix element of the renormalized operator 
$O_+^{\rm ren}(\omega,\mu)$ is, up to a Dirac trace, given by the product 
$F(\mu)\,\phi_+^B(\omega,\mu)$. It follows that the LCDA obeys the 
evolution equation
\begin{equation}\label{phievol}
   \frac{d}{d\ln\mu}\,\phi_+^B(\omega,\mu)
   = - \int_0^\infty\!d\omega'\,\gamma_+(\omega,\omega',\mu)\,
   \phi_+^B(\omega',\mu)
\end{equation}
with the anomalous dimension (unless otherwise indicated, 
$\alpha_s\equiv\alpha_s(\mu)$)
\begin{eqnarray}\label{gampl}
   \gamma_+(\omega,\omega',\mu)
   &=& - \int d\tilde\omega\,
    \frac{dZ_+(\omega,\tilde\omega,\mu)}{d\ln\mu}\,
    Z_+^{-1}(\tilde\omega,\omega',\mu) \nonumber\\
   &&-\,\gamma_F(\alpha_s)\,\delta(\omega-\omega') \,.
\end{eqnarray}
Here $\gamma_F$ is the universal anomalous dimension of local heavy-light 
currents in HQET, which determines the scale dependence of $F(\mu)$. 

\begin{figure}
\epsfxsize=7.0cm
\centerline{\epsffile{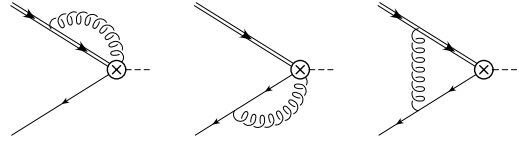}}
\vspace{0.2cm}
\centerline{\parbox{14cm}{\caption{\label{fig:SCET}
One-loop diagrams contributing to the calculation of the anomalous 
dimension. The crossed circle denotes an insertion of the operator 
$O_+^{\rm bare}(\omega')$. The double lines represent heavy-quark fields 
in HQET.}}}
\end{figure}

At one-loop order the result for $Z_+(\omega,\omega',\mu)$ is obtained by 
evaluating the UV poles of the one-loop HQET diagrams shown in 
Figure~\ref{fig:SCET}. The loop integrals contain $\delta$-functions 
whose arguments depend on the value of the ``plus component'' 
$k_+=n\cdot k$ of the loop momentum. The integration over the orthogonal 
``minus component'' is most easily performed using the theorem of 
residues, and the remaining integration over transverse components can be 
evaluated using standard techniques of dimensional regularization 
(setting $d_\perp=2-2\epsilon$ for the dimension of the transverse 
space). In the first two diagrams the transverse loop integrations give 
rise to UV poles, whereas the third diagram is UV finite. In units of 
$C_F\alpha_s/4\pi$, the corresponding contributions to 
$Z_+(\omega,\omega',\mu)$ are given by ($k\equiv -k_+$)
\begin{eqnarray}\label{diags}
   D_1 &=& - \frac{2}{\epsilon} \int_0^\infty \frac{dk}{k}
    \left( \frac{k}{\mu} \right)^{-2\epsilon}
    \Big[ \delta(k+\omega'-\omega) - \delta(\omega-\omega') \Big] \,,
    \nonumber\\[-0.1cm]
   && \\[-0.3cm]
   D_2 &=& - \frac{2}{\epsilon} \int_0^{\omega'} \frac{dk}{k}\,
    \frac{\omega'-k}{\omega'}\,
    \Big[ \delta(k+\omega-\omega') - \delta(\omega-\omega') \Big] \,.
    \nonumber
\end{eqnarray}
These expressions must be understood in the sense of distributions, and
only pole terms in $1/\epsilon$ are kept after integration over $\omega'$ 
in (\ref{Zdef}). Note that the $k$-integral in the first diagram in 
Figure~\ref{fig:SCET} gives rise to a $1/\epsilon^2$ pole times 
$\delta(\omega-\omega')$. Accounting for wave-function renormalization of 
the external quark fields we obtain
\begin{eqnarray}\label{Zpl}
   Z_+^{(1)}(\omega,\omega',\mu)
   &=& \left( \frac{1}{\epsilon^2}
    + \frac{2}{\epsilon}\,\ln\frac{\mu}{\omega} - \frac{5}{2\epsilon}
    \right) \delta(\omega-\omega') \nonumber\\
   &-& \frac{2}{\epsilon}\,\left[ \frac{\omega}{\omega'}\,
    \frac{\theta(\omega'-\omega)}{\omega'-\omega}
    + \frac{\theta(\omega-\omega')}{\omega-\omega'} \right]_+ .
\end{eqnarray}
Here and below a superscript ``(1)'' is used to indicate one-loop 
coefficients in units of $C_F\alpha_s/4\pi$. The plus distribution is 
defined such that, when $Z_+$ is integrated with a function 
$f(\omega')$, one must replace $f(\omega')\to f(\omega')-f(\omega)$ 
under the integral. The non-diagonal terms in (\ref{Zpl}) agree with 
those found in \cite{Grozin:1996pq}; however, the double pole and 
logarithm in the first term were missed in that paper. For the anomalous 
dimension in (\ref{gampl}) we now obtain
\begin{eqnarray}\label{me}
   \gamma_+(\omega,\omega',\mu)
   &=& \left[ \Gamma_{\rm cusp}(\alpha_s)\,\ln\frac{\mu}{\omega}
    + \gamma(\alpha_s) \right]\,\delta(\omega-\omega') \nonumber\\
   &&\mbox{}+ \omega\,\Gamma(\omega,\omega',\alpha_s) \,,
\end{eqnarray}
where $\int d\omega'\,\Gamma(\omega,\omega',\alpha_s)=0$ by definition. 
Taking into account that $\gamma_F^{(1)}=-3$ is the one-loop coefficient 
of the anomalous dimension of heavy-light currents, it follows from 
(\ref{Zpl}) that $\Gamma_{\rm cusp}^{(1)}=4$, $\gamma^{(1)}=-2$, and 
\begin{equation}\label{G1l}
   \Gamma^{(1)}(\omega,\omega')
   = - \Gamma_{\rm cusp}^{(1)} \left[ 
   \frac{\theta(\omega'-\omega)}{\omega'(\omega'-\omega)}
   + \frac{\theta(\omega-\omega')}{\omega(\omega-\omega')} \right]_+ .~
\end{equation}

The anomalous dimension in (\ref{me}) is of the Sudakov type. It 
contains a logarithmic dependence on the renormalization scale in 
addition to the dependence through the running coupling 
$\alpha_s(\mu)$. This feature distinguishes the kernel for the $B$-meson
LCDA from the familiar Brodsky--Lepage kernel for the LCDA of a light
pseudoscalar meson \cite{Lepage:1980fj}. The extra logarithm has its 
origin in the renormalization properties of Wilson lines with light-like 
segments \cite{Korchemsky:wg}. Using the fact that a heavy quark in HQET 
can be described by a Wilson line $h(0)=S_v(0,-\infty)\,h_0$, where $h_0$ 
is a sterile field without QCD interactions, it follows that the 
interacting fields in the operator in (\ref{LCDA}) can be written as 
$\bar q_s(z)\,S_n(z,0)\,S_v(0,-\infty)$. This corresponds to a gauge 
string extending from minus infinity to 0 along the $v$ direction, 
another string extending from 0 to $z$ along the light-like $n$ 
direction, and a light quark field located at the end of the string at 
position $z$. This Wilson line has a cusp singularity at the origin, 
which gives rise to a local, single-logarithmic term in the anomalous 
dimension. From (\ref{diags}) it is seen that the logarithm arises indeed 
from the gluon exchange between the two strings $S_n$ and $S_v$ (first 
diagram in Figure~\ref{fig:SCET}). Such cusp singularities are universal 
and of geometric origin. The corresponding anomalous dimension 
$\Gamma_{\rm cusp}$ is process independent and has been computed at 
two-loop order in \cite{Korchemsky:wg}. The important observation 
following from this discussion is that also in higher orders there is 
only a single logarithm present in the anomalous dimension
$\gamma_+(\omega,\omega',\mu)$. With this knowledge, the evolution 
equation can be solved in RG-improved perturbation theory.

From the scale independence of the convolution integral (\ref{conv}) it 
follows that the hard-scattering kernel obeys the evolution equation
\begin{equation}\label{Tevol}
   \frac{d}{d\ln\mu}\,T(\omega,\mu)
   = \int_0^\infty\!d\omega'\,\frac{\omega}{\omega'}\,
   \gamma_+(\omega',\omega,\mu)\,T(\omega',\mu) \,.
\end{equation}
The explicit result for the anomalous dimension given above shows that 
$\frac{\omega}{\omega'}\,\gamma_+(\omega',\omega,\mu)
=\gamma_+(\omega,\omega',\mu)$ (at least to one-loop order). Hence, apart 
from an overall sign the LCDA and the kernel obey the same RG equation.

\vspace{0.3cm}
{\em 3. Analytic solutions.}\quad
The general solution of the evolution equations (\ref{phievol}) and 
(\ref{Tevol}) can be obtained using the fact that on dimensional grounds
\begin{equation}\label{Fdef}
   \int d\omega'\,\omega\,\Gamma(\omega,\omega',\alpha_s)\,(\omega')^a
   = \omega^a\,\F(a,\alpha_s) \,,
\end{equation}
where the function $\F$ only depends on the exponent $a$ and the coupling 
constant, and $\F(0,\alpha_s)=0$ by definition. The integral on the 
left-hand side is convergent as long as $-1<\mbox{Re}\,a<1$. The 
corresponding integral with $\Gamma(\omega,\omega',\alpha_s)$ replaced by 
$\Gamma(\omega',\omega,\alpha_s)$, which is relevant to the evolution 
equation (\ref{Tevol}) for the kernel, is given by $\F(-a,\alpha_s)$. 
This follows from the fact that 
$\Gamma(\omega,\omega',\alpha_s)=\omega^{-2}\,f(\omega'/\omega)$ on 
dimensional grounds. At one-loop order we find from (\ref{G1l})
\begin{equation}\label{F1loop}
   \F^{(1)}(a) = \Gamma_{\rm cusp}^{(1)}\,
   \Big[ \psi(1+a) + \psi(1-a) + 2\gamma_E \Big] \,,
\end{equation}
where $\psi(z)$ is the logarithmic derivative of the Euler 
$\Gamma$-function. 

Relation (\ref{Fdef}) implies that the ansatz 
\begin{equation}\label{ansatz}
   f(\omega,\mu,\mu_0,\eta) = \left( \frac{\omega}{\mu_0}
   \right)^{\eta+g(\mu,\mu_0)} \exp U(\mu,\mu_0,\eta)
\end{equation}
with
\begin{eqnarray}\label{myeqs}
   g(\mu,\mu_0)
   &=& \int\limits_{\alpha_s(\mu_0)}^{\alpha_s(\mu)}\!
    d\alpha\,\frac{\Gamma_{\rm cusp}(\alpha)}{\beta(\alpha)} \,,
    \nonumber\\[-0.2cm]
   && \\[-0.2cm]
   U(\mu,\mu_0,\eta)
   &=& - \int\limits_{\alpha_s(\mu_0)}^{\alpha_s(\mu)}\!
    \frac{d\alpha}{\beta(\alpha)}\,\Big[
    g(\mu,\mu_\alpha) + \gamma(\alpha) \nonumber\\
   &&\qquad\mbox{}+ 
    \F\big(\eta + g(\mu_\alpha,\mu_0),\alpha \big) \Big] \,, \nonumber
\end{eqnarray}
provides a solution to the evolution equation (\ref{phievol}) with 
initial condition $f(\omega,\mu_0,\mu_0,\eta)=(\omega/\mu_0)^\eta$ at 
some scale $\mu_0$. Here $\mu_\alpha$ is defined such that 
$\alpha_s(\mu_\alpha)=\alpha$, the $\beta$-function is 
$\beta(\alpha_s)=d\alpha_s/d\ln\mu$, and $\eta$ can be an arbitrary 
complex parameter. Note that $g(\mu,\mu_0)>0$ if $\mu>\mu_0$. Given the 
exact expressions in (\ref{myeqs}) one can derive approximate results for 
the functions $g$ and $U$ at a given order in RG-improved perturbation 
theory. The explicit forms arising at next-to-leading order can be found 
in \cite{Bosch:2003fc}.

We now assume that the function $\phi_+^B(\omega,\mu_0)$ is given at some 
low scale $\mu_0\sim\mbox{few}\times\Lambda_{\rm QCD}$ and define its 
Fourier transform with respect to $\ln(\omega/\mu_0)$ through
\begin{equation}\label{FT}
   \phi_+^B(\omega,\mu_0)
   = \frac{1}{2\pi} \int_{-\infty}^\infty\!dt\,\varphi_0(t)
   \left( \frac{\omega}{\mu_0} \right)^{it} \,,
\end{equation}
where $\varphi_0(0)=1/\lambda_B$ is determined in terms of the first 
inverse moment of the LCDA at the scale $\mu_0$ \cite{Beneke:1999br}. It 
follows that the result for the LCDA at a different scale $\mu$ is given 
by
\begin{equation}\label{phisol}
   \phi_+^B(\omega,\mu)
   = \frac{1}{2\pi} \int_{-\infty}^\infty\!dt\,\varphi_0(t)\,
   f(\omega,\mu,\mu_0,it) \,. 
\end{equation}
It is instructive to work out the $t$ and $\omega$-dependence of the 
integrand at leading order in RG-improved perturbation theory. Using the 
one-loop result (\ref{F1loop}) we obtain
\begin{equation}
   f(\omega,\mu,\mu_0,it)\propto
   \left( \frac{\omega}{\mu_0} \right)^{it+g} 
   \frac{\Gamma(1-it-g)\,\Gamma(1+it)}{\Gamma(1+it+g)\,\Gamma(1-it)} \,,
\end{equation}
where $g=(2C_F/\beta_0)\,\ln[\alpha_s(\mu_0)/\alpha_s(\mu)]$ is the 
leading-order contribution to the function $g(\mu,\mu_0)$, and the 
formula can only be trusted if $g<1$ (which is justified for all 
reasonable parameter values). The expression above is an analytic 
function in the complex $t$-plane with singularities along the imaginary 
axis. For $\mu>\mu_0$ the nearest singularities are located at 
$t=-i(1-g)$ and $t=i$. The locations of the nearest singularities in the 
product $\varphi_0(t)\,f(\omega,\mu,\mu_0,it)$ determine the asymptotic 
behavior of the LCDA for $\omega\to 0$ (lower half-plane) and 
$\omega\to\infty$ (upper half-plane). If we assume that the function 
$\phi_+^B(\omega,\mu_0)$ at the low scale vanishes like $\omega^\delta$ 
for small $\omega$ and falls off like $\omega^{-\xi}$ for large $\omega$ 
(exponential fall-off would correspond to $\xi\to\infty$), then the 
nearest poles in $\varphi_0(t)$ are located at $t=-i\delta$ and $t=i\xi$, 
respectively. It follows that after evolution effects
\begin{equation}
   \phi_+^B(\omega,\mu)\sim
   \cases{ \omega^{\min(1,\delta+g)} \,; & for $\omega\to 0$, \cr
           \omega^{-\min(1,\xi)+g} \,; & for $\omega\to\infty$. \cr}
\end{equation}
Irrespective of the initial behavior of the LCDA, evolution effects drive 
it toward a linear growth at the origin and generate a radiative tail 
that falls off slower than $1/\omega$ even if the initial function has an 
arbitrarily rapid fall-off. This implies, in particular, that the 
normalization integral of $\phi_+^B(\omega,\mu)$ is UV divergent for 
large values of $\omega$. This fact, which was already noted in 
\cite{Grozin:1996pq}, is not an obstacle to our analysis.
Convolution integrals appearing in QCD factorization theorems are always 
of the form (\ref{conv}), in which the extra $1/\omega$ factor suppresses 
the contributions from large $\omega$ and renders the integral finite.  

The solution of the evolution equation (\ref{Tevol}) for the 
hard-scattering kernel proceeds in a similar way, except that in the 
expression for the function $f$ in (\ref{ansatz}) one must replace 
$\mu_0\to\mu_i$, $g(\mu,\mu_0)\to-g(\mu,\mu_i)$, and
$U(\mu,\mu_0,\eta)\to -U(\mu,\mu_i,-\eta)$. As mentioned in the 
Introduction, at fixed order in perturbation theory and at an 
intermediate scale $\mu_i\sim\sqrt{m_b\Lambda}$, the kernel depends on 
$\omega$ only through logarithms of the type $\ln(2E\omega/\mu_i^2)$, 
which are of order 1 and so do not need to be resummed. We can therefore 
write $T(\omega,\mu_i)\equiv\T[\ln(2E\omega/\mu_i^2),\dots]$, where the 
dots represent other arguments independent of $\omega$. (Some of these 
arguments may contain large logarithms, which must be resummed 
separately.) It follows that by considering derivatives of the solution 
with respect to $\eta$ (evaluated at $\eta=0$) we can satisfy arbitrary 
initial conditions at the scale $\mu_i$. We can then solve (\ref{Tevol}) 
and compute the hard-scattering kernel at a scale $\mu<\mu_i$. The exact 
solution is given by
\begin{eqnarray}\label{magic1}
   T(\omega,\mu)
   &=& \T[\nabla_\eta,\dots] \left( \frac{2E\omega}{\mu_i^2}
    \right)^\eta \left( \frac{\omega}{\mu_i}
    \right)^{g(\mu_i,\mu)} \nonumber\\
   &&\mbox{}\times\exp[-U(\mu,\mu_i,-\eta)] \Big|_{\eta=0} \,,
\end{eqnarray}
where the notation $\T[\nabla_\eta,\dots]$ means that one must replace 
each logarithm of the ratio $2E\omega/\mu_i^2$ in the initial condition 
by a derivative with respect to the auxiliary parameter $\eta$. 
It follows that the kernel scales like $T\sim\omega^{g(\mu_i,\mu)}$
modulo logarithms.

From the explicit solutions given above it can be seen that the 
convolution integral in (\ref{conv}) is indeed independent of the
renormalization scale. For this, it is necessary to move the
integration contour in (\ref{phisol}) into the upper complex $t$ plane 
by an amount $\eta+g(\mu_i,\mu_0)$. Note, however, that the product 
$T(\omega,\mu)\,\phi_+^B(\omega,\mu)$ for fixed $\omega$ is not scale 
independent. We also stress that evolution effects mix different 
(fractional) moments of the LCDA. For instance, the first inverse moment 
of the LCDA at a scale $\mu$, which is sometimes called 
$1/\lambda_B(\mu)$, is related to a fractional inverse moment of order 
$1-g(\mu,\mu')$ at a different scale $\mu'$. As a result, the scale 
dependence of the parameter $\lambda_B(\mu)$ is not calculable in 
perturbation theory. Controlling it would require knowledge of the 
functional form of the LCDA.

\begin{figure}
\epsfxsize=8.6cm
\centerline{\epsffile{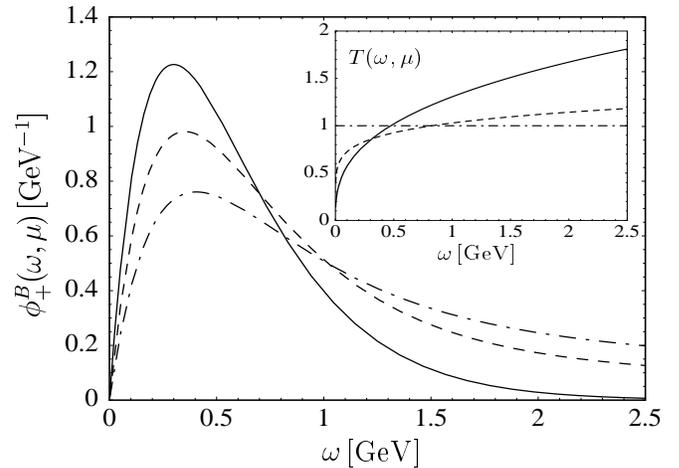}}
\vspace{0.2cm}
\centerline{\parbox{14cm}{\caption{\label{fig:LCDA}
Results for the LCDA and the kernel (inset) for different values of the 
renormalization scale such that $\alpha_s=1$ (solid), 0.5 (dashed), 0.3 
(dashed-dotted). We use $\lambda_B=0.3$\,GeV at $\mu=\mu_0$.}}}
\end{figure}

To illustrate our results we consider a scenario where the 
hard-scattering kernel at a high scale $\mu_i$ with $\alpha_s(\mu_i)=0.3$ 
takes the simple form $T(\omega,\mu_i)=1$, and where the LCDA at a low 
hadronic scale $\mu_0$ with $\alpha_s(\mu_0)=1$ assumes the form
$\phi_+^B(\omega,\mu_0)=(\omega/\lambda_B^2)\,e^{-\omega/\lambda_B}$
suggested in \cite{Grozin:1996pq}, for which 
\begin{equation}
   \varphi_0(t) = \frac{1}{\lambda_B}
   \left( \frac{\mu_0}{\lambda_B} \right)^{it} \Gamma(1-it) \,.
\end{equation}
Figure~\ref{fig:LCDA} shows the results for the LCDA and the kernel at 
three different values of the renormalization scale. For simplicity, the
anomalous dimensions and $\beta$-function are evaluated at one-loop 
order. Whereas the kernel exhibits a smooth power-like behavior, the most 
characteristic feature of the evolution of the LCDA is the development of 
a radiative tail for large values of $\omega$. In this example the value 
of the convolution integral (\ref{conv}) is about 25\% smaller than the 
value $1/\lambda_B$ which one would obtain without evolution effects.

\vspace{0.3cm}
{\em 4. Conclusions.}\quad
We have derived evolution equations for the leading-order $B$-meson LCDA 
and for the hard-scattering kernels entering the spectator term in QCD 
factorization theorems for exclusive $B$ decays into light particles. 
Simple scaling relations are obtained for the asymptotic behavior of 
these quantities. Exact analytic solutions to the evolution equations are 
given in terms of integrals over anomalous dimension functions. This 
accomplishes the resummation of large Sudakov double logarithms to all 
orders in perturbation theory.

\vspace{0.3cm}  
{\it Acknowledgment:\/} 
We are grateful to Richard Hill for useful discussions. This research was 
supported by the National Science Foundation under Grant PHY-0098631.

\end{document}